\documentclass[doublecol]{epl2} 
\usepackage{epsfig}
\usepackage{bm}

\title{{\it Europhysics Letters, 85 (2009) 49002}\\
~~~~~\\
Transitional dynamics of the solar convection zone}
\shorttitle{Transitional dynamics of the solar convection zone} %Insert here a short version of the title if it exceeds 70 characters

\author{ A. Bershadskii\inst{1,2}}
\shortauthor{A. Bershadskii}

\institute{                    
  \inst{1} ICAR, P.O. Box 31155, Jerusalem 91000, Israel\\
  \inst{2} ICTP, Strada Costiera 11, I-34100 Trieste, Italy}


\abstract{Solar activity is studied using a cluster analysis of the time-fluctuations 
of sunspot number. In the historic period
(1850-1932) the cluster exponent $\alpha \simeq 0.37$ (strong
clustering) for the high activity components of the solar cycles.
In the modern period (last seven solar cycles: 1933-2007) the
cluster exponent was $\alpha \simeq 0.50$ (random, white noise-like). 
Comparing these results with the corresponding
data from laboratory experiments on convection it is shown,
that in the historic period emergence of sunspots in the solar
photosphere was dominated by turbulent photospheric
convection. In the modern period, this domination was broken by a
new more active dynamics of the inner layers of the convection zone. 
Cluster properties of the solar wind magnetic field and the 
aa-geomagnetic-index also support this result. 
Long-range chaotic dynamics in the solar activity is briefly discussed.}

\begin{document}

\maketitle

\section{Introduction} 

 The sunspot number is the main direct and
reliable source of information about the solar dynamics for a historic
period. In a recent paper \cite{sol}, for instance, results of a
reconstruction (based on radiocarbon concentrations, see also
\cite{usos1}) of the sunspot number were presented for the past
11,400 years. Analysis of this reconstruction showed an exceptional level of solar
activity during the past seven solar cycles (1933-2007).
%%%%%%% FIGURE 1 %%%%%%%%%%%%%%%%%%
\begin{figure} \vspace{+0.3cm}\centering
\epsfig{width=.45\textwidth,file=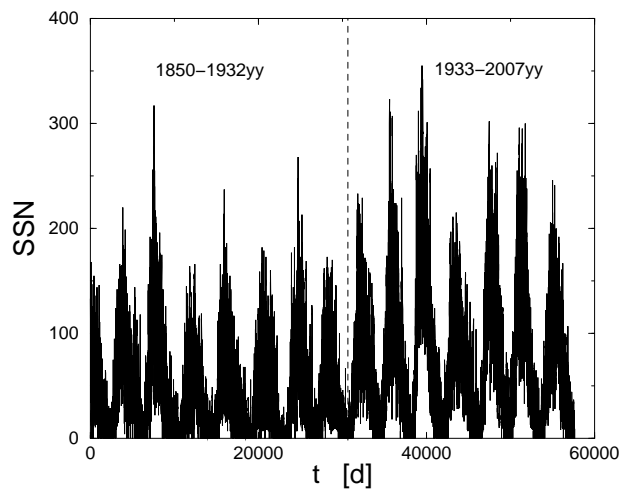} \vspace{-0cm}
\caption{Daily sunspot number (SSN) vs time \cite{belg}. The dashed
straight line separates between historic and modern periods. }
\end{figure}
%%%%%%%%%%%%%%%%%%%%%%%%%%%%%%%%%%%
Sunspots appear as the visible counterparts of
magnetic flux tubes in the convective zone of the sun. Since a
strong magnetic field is considered as a primary phenomenon that
controls generation of the sunspots the crucial question is: Where
has the magnetic field itself been generated? The location of the
solar dynamo is the subject of vigorous discussions in recent years. A
general consensus had been developed to consider
the shear layer at the {\it bottom} of the convection zone as the main
source of the solar magnetic field \cite{sw} (see, for a recent
review \cite{brand1}). In recent years, however, the existence of a
prominent radial shear layer near the top of the convection zone has
become rather obvious and the problem again became actual. The presence
of large-scale meandering flow fields (like jet streams), banded
zonal flows and evolving meridional circulations together with
intensive multiscale turbulence shows that the near surface layer is
a very complex system, which can significantly affect the processes
of the magnetic field and the sunspots generation. There could be 
two sources for the poloidal magnetic field: one near the
bottom of the convection zone (or just below it \cite{sw}), another 
one resulting from an active-region tilt near the surface of the
convection zone. For the recently renewed Babcock-Leighton
\cite{bab},\cite{leig} solar dynamo scenario, for instance, a
combination of the sources was assumed for predicting future solar
activity levels \cite{dtg}, \cite{ccj}. In this scenario the surface
generated poloidal magnetic field is carried to the bottom of the
convection zone by turbulent diffusion or by the meridional
circulation. The toroidal magnetic
field is produced from this poloidal field by differential rotation in the bottom shear layer.
Destabilization and emergence of the toroidal fields (in the form of
curved tubes) due to magnetic buoyancy can be considered as a source
of the pairs of sunspots of opposite polarity. The turbulent
convection in the convection zone and, especially, in the
near-surface layer captures the magnetic flux tubes and either {\it
disperses} or {\it pulls} them trough the surface to become
sunspots.
%%%%%%% FIGURE 2 %%%%%%%%%%%%%%%%%%
\begin{figure} \vspace{-0.5cm}\centering
\epsfig{width=.45\textwidth,file=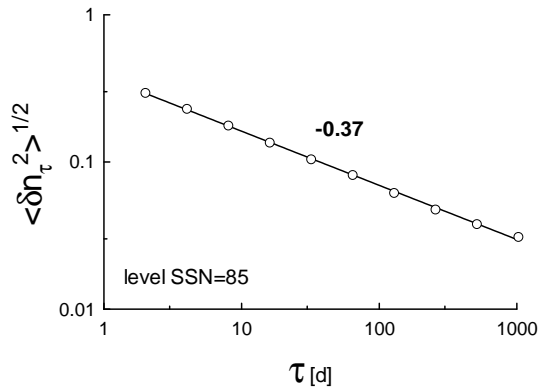} \vspace{-4cm}
\caption{The standard deviation for $\delta n_{\tau}$ vs $\tau$ for
the level SSN=85 (historic period) in log-log scales. The straight
line (the best fit) indicates the scaling law Eq. (1). }
\end{figure}
%%%%%%%%%%%%%%%%%%%%%%%%%%%%%%%%%%%
%%%%%%% FIGURE 3 %%%%%%%%%%%%%%%%%%
\begin{figure} \vspace{-0.5cm}\centering
\epsfig{width=.45\textwidth,file=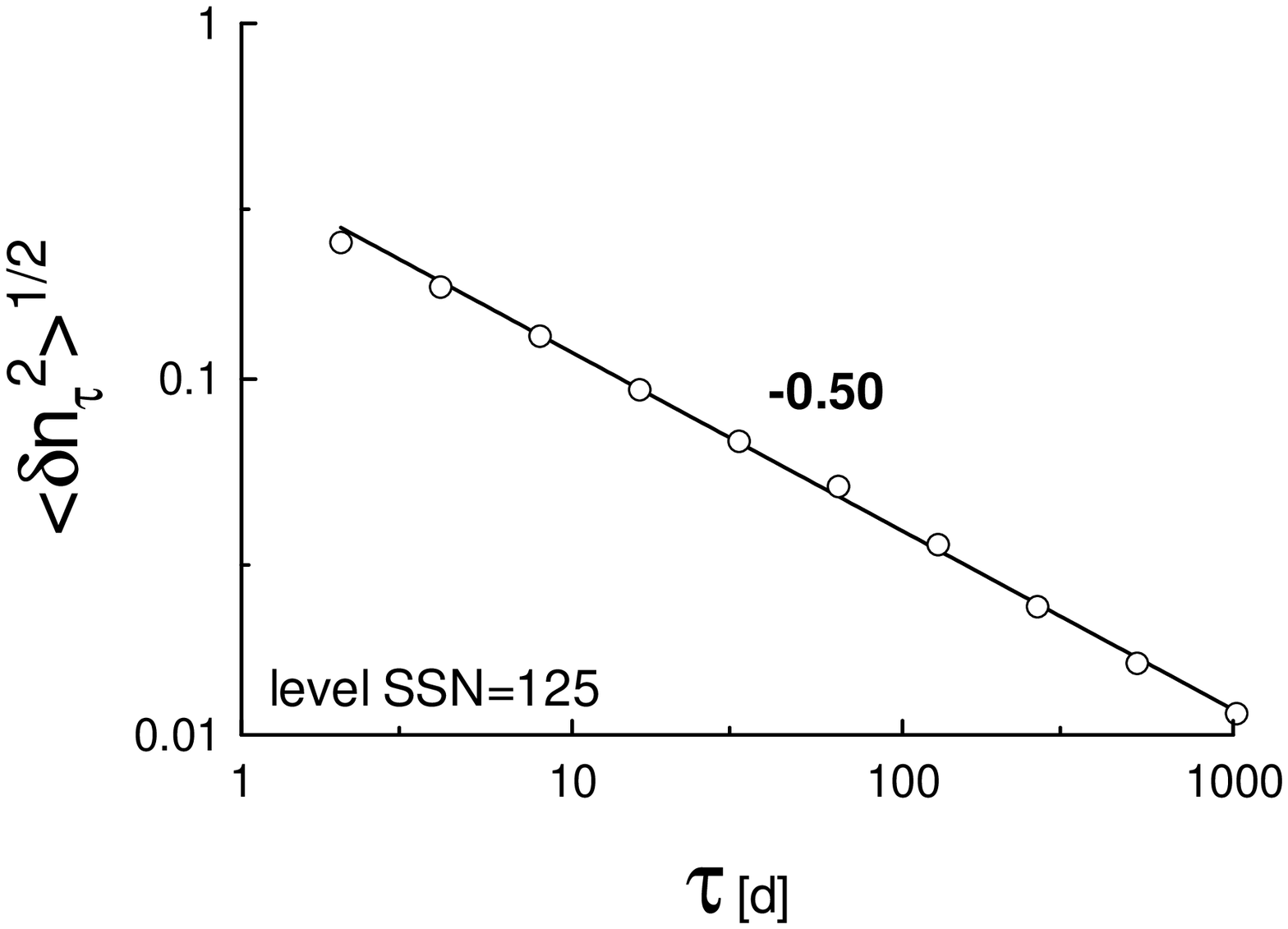} \vspace{-4cm}
\caption{The standard deviation for $\delta n_{\tau}$ vs $\tau$ for
the level SSN=125 (modern period) in log-log scales. The straight
line (the best fit) indicates the scaling law Eq. (1).  }
\end{figure}
%%%%%%%%%%%%%%%%%%%%%%%%%%%%%%%%%%%
%%%%%%% FIGURE 4%%%%%%%%%%%%%%%%%%
\begin{figure} \vspace{-0.5cm}\centering
\epsfig{width=.45\textwidth,file=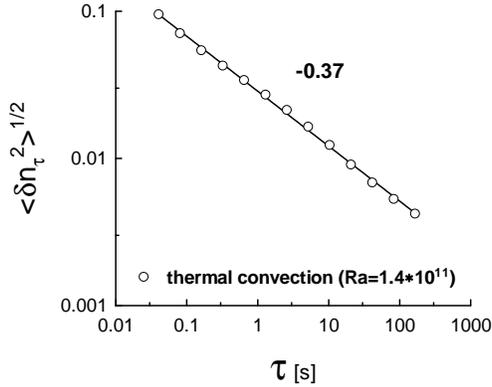} \vspace{-4cm}
\caption{The standard deviation for $\delta n_{\tau}$ vs $\tau$ for
the temperature fluctuations in the Rayleigh-Bernard convection
laboratory experiment with $Ra=1.4 \times 10^{11}$ \cite{ns}. The
straight line indicates the scaling law Eq. (1). }
\end{figure}
%%%%%%%%%%%%%%%%%%%%%%%%%%%%%%%%%%%

The magnetic field plays a passive role in the photosphere
and does not participate significantly in the turbulent photospheric 
energy transfer. On the other hand, the very complex and turbulent
near-surface layer (including photosphere) can significantly affect
the process of emergence of sunspot. The
similarity of light elements properties in the spot umbra and
granulation is one of the indication of such phenomena.

The paper reports a direct relation between the fluctuations in sunspot number 
and the temperature of turbulent convection in the photosphere for the historic period. 
This relation allows for certain conclusions
about the generation mechanisms of the magnetic fields and the
sunspots. In the modern period (last seven solar cycles: 1933-2007, cf.
\cite{sol},\cite{usos1}) the relative role of the surface layer
(photosphere) in the process of emergence of sunspots decreased
in comparison with the historic period (the seven solar cycles
preceding 1933), implying a drastic increase of the relative role
of the inner layers of the convection zone
in the modern period.

\section{Time-clustering of fluctuations}

 In order to extract a new
information from sunspot number data we apply the
fluctuation clustering analysis suggested in the Ref. \cite{sb}. For 
a time depending signal we count the number of
'zero'-crossings of the signal (the points on the time axis
where the signal is equal to zero) in a time interval $\tau$ and
consider their running density $n_{\tau}$. Let us denote
fluctuations of the running density as $\delta n_{\tau} = n_{\tau} -
\langle n_{\tau} \rangle$, where the brackets mean the average over
long times. We are interested in scaling variation of the standard
deviation of the running density fluctuations $\langle \delta
n_{\tau}^2 \rangle^{1/2}$ with $\tau$
$$
\langle \delta n_{\tau}^2 \rangle^{1/2} \sim \tau^{-\alpha}
\eqno{(1)}
$$
For white noise signal it can be derived analytically
\cite{molchan},\cite{lg} that $\alpha = 1/2$ (see also \cite{sb}).
The same consideration can be applied not only to the
'zero'-crossing points but also to any level-crossing points of the
signal.

In Fig. 1 we show the daily sunspot number (SSN) for the period
1850-2007 \cite{belg}. Even by eye one can see that the modern
period (last seven cycles, 1933-2007) is different from the
corresponding historic period: 1850-1933. Therefore, in order to
calculate the cluster exponent (if exists) for this signal one
should make this calculation separately for the modern and for the
historic periods. We are interested in the active parts of the solar
cycles. Therefore, for the historic period let us start from the
level SSN=85. The set of the level-crossing points has a few large
voids corresponding to the weak activity periods. To make the set
statistically stationary we will cut off these voids. The remaining
set (about $10^4$ data-points) exhibits good statistical stationarity 
that allows to calculate
scaling exponents corresponding to this set. Fig. 2 shows (in the
log-log scales) dependence of the standard deviation of the running
density fluctuations $\langle \delta n_{\tau}^2 \rangle^{1/2}$ on
$\tau$. The straight line is drawn in this figure to indicate the
scaling (1). The slope of this straight line provides us with the
cluster-exponent  $\alpha = 0.37 \pm 0.02$. This value turned to be
insensitive to a reasonable variation of the SSN level. Results of
analogous calculations performed for the modern period are shown in
Fig. 3 for the SSN level SSN=125. The calculations performed for the
modern period provide us with the cluster-exponent $\alpha=0.5 \pm
0.02$ (and again this value turned to be insensitive to a reasonable
variation of the SSN level).

The exponent $\alpha\simeq 0.5$ (for the modern period) indicates a
random (white noise like) situation. While the exponent
$\alpha\simeq 0.37$ (for the historic period) indicates strong
clustering. The question is: Where is this strong clustering coming
from? It is shown in the paper \cite{sb} that signal produced by
turbulence exhibit strong clustering. Moreover, the cluster
exponents for these signals depend on the turbulence intensity and
they are nonsensitive to the types of the boundary conditions.
Fortunately, we have direct estimates of the value of the main
parameter characterizing intensity of the turbulent convection in
photosphere: Rayleigh number $Ra \sim 10^{11}$ (see, for instance
\cite{bld}). In Fig. 4 we show calculation of the cluster exponent
for the temperature fluctuations in the classic Rayleigh-Bernard
convection laboratory experiment for $Ra \sim 10^{11}$ (for a
description of the experiment details see \cite{ns}). The calculated
value of the cluster exponent $\alpha =0.37\pm 0.01$ coincides with
the value of the cluster-exponent obtained above for the sunspot
number fluctuations for the historic period. If the value of the
Rayleigh number $Ra $ in the photosphere for the historic period has the 
same order as for the modern period: $Ra \sim10^{11}$ (see next Section), 
then one can suggest that the
clustering of the sunspot number fluctuations in the historic period
is due to strong modulation of these fluctuations by the turbulent
fluctuations of the temperature in the photospheric convection. This
seems to be natural for the case when the photospheric convection
{\it determines} the sunspot emergence in the photosphere. However,
in the case when the effect of the photospheric convection on the
SSN fluctuations is comparable with the effects of the inner
convection zone layers on the SSN fluctuations the clustering should
be randomized by the mixing of the sources, and the cluster exponent
$\alpha \simeq 0.5$ (similar to the white noise signal). The last
case apparently takes place for the modern period.
%%%%%%% FIGURE 5 %%%%%%%%%%%%%%%%%%
\begin{figure} \vspace{-0.5cm}\centering
\epsfig{width=.45\textwidth,file=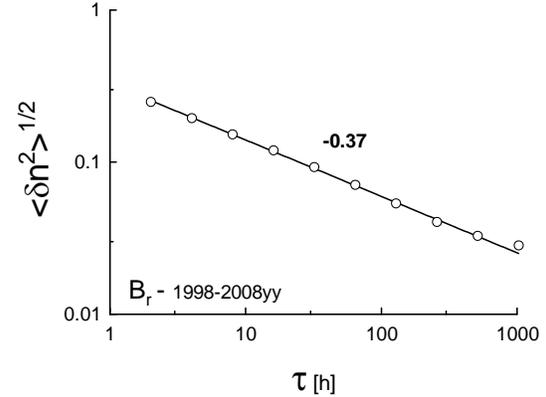} \vspace{-5cm} \caption{
The standard deviation for $\delta n_{\tau}$ vs. $\tau$ 
for radial component $B_r$ 
of the interplanetary magnetic field, as measured by
the ACE magnetometers for the last solar cycle (hourly average \cite{ACE}). 
}
\end{figure}
%%%%%%%%%%%%%%%%%%%%%%%%%%%%%%%%%%%
%%%%%%% FIGURE 6 %%%%%%%%%%%%%%%%%%
\begin{figure} \vspace{-0.5cm}\centering
\epsfig{width=.45\textwidth,file=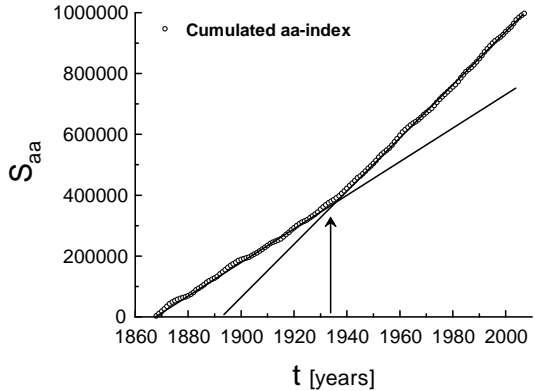} \vspace{-4.5cm} \caption{
Cumulative aa-index vs. time. Daily aa-index was taken from \cite{wdc}. 
The arrow indicates beginning of the transitional solar cycle.}
\end{figure}
%%%%%%%%%%%%%%%%%%%%%%%%%%%%%%%%%%%
%%%%%%% FIGURE 7 %%%%%%%%%%%%%%%%%%
\begin{figure} \vspace{-0.5cm}\centering
\epsfig{width=.45\textwidth,file=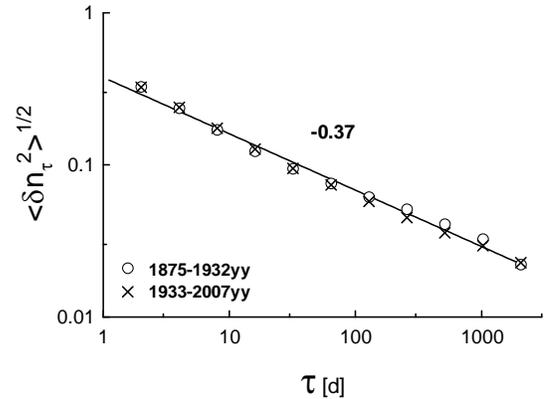} \vspace{-4.5cm} \caption{
The standard deviation for $\delta n_{\tau}$ vs. $\tau$ (in log-log scales) 
for the daily aa-index \cite{wdc}: 
circles correspond to the historic period and crosses correspond to the modern period. }
\end{figure}
%%%%%%%%%%%%%%%%%%%%%%%%%%%%%%%%%%%
%%%%%%% FIGURE 8 %%%%%%%%%%%%%%%%%%
\begin{figure} \vspace{-0.5cm}\centering
\epsfig{width=.45\textwidth,file=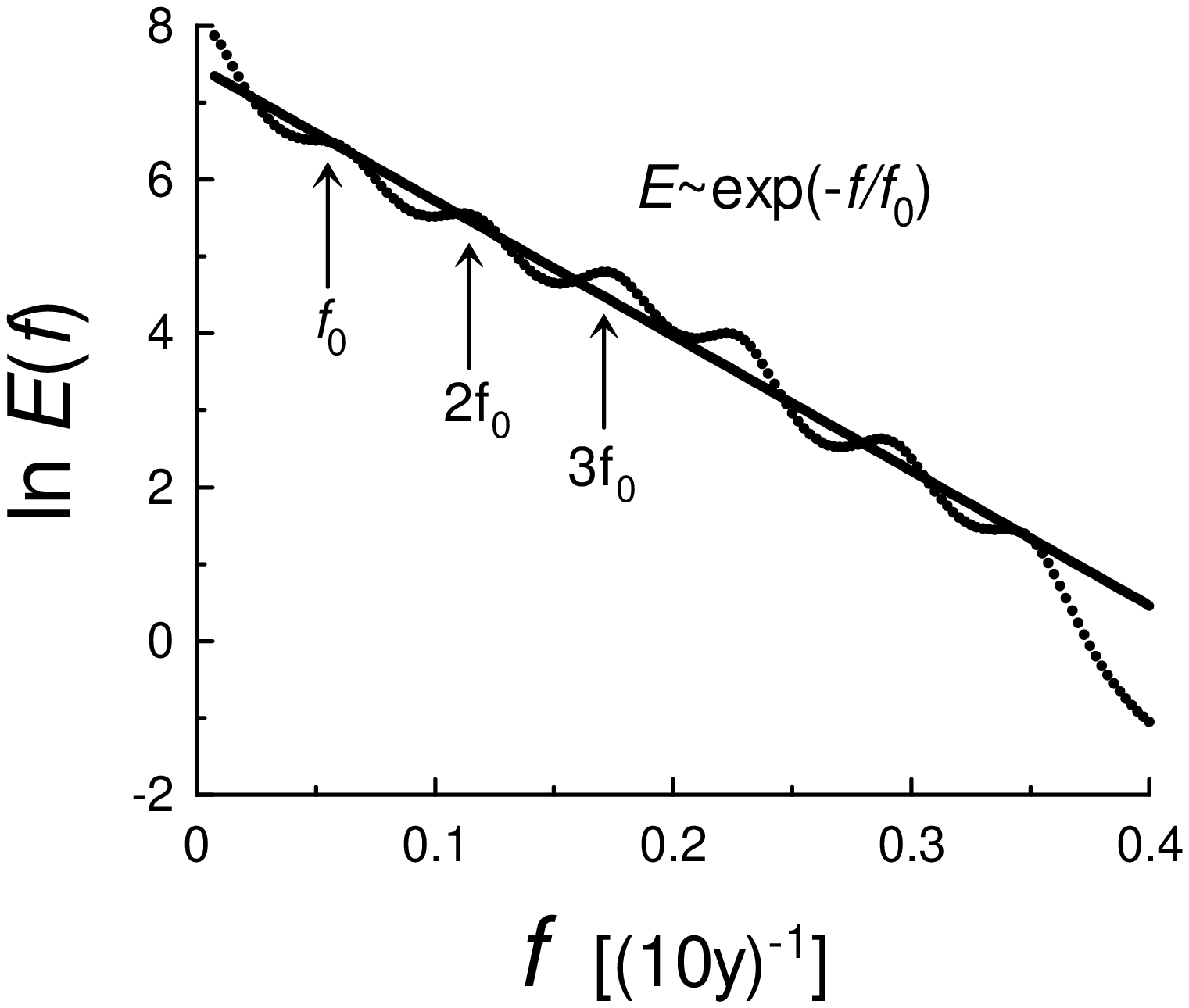} \vspace{-4.5cm}
\caption{Spectrum of the sunspots number fluctuations in the ln-linear scales (the reconstructed data 
for the last 11,000 years have been taken from \cite{usos-recon}). The straight line is drawn to indicate 
the exponential law Eq. (3).}
\end{figure}
%%%%%%%%%%%%%%%%%%%%%%%%%%%%%%%%%%%
Since the Rayleigh number $Ra$ of the photospheric convection preserves 
its order $Ra \sim 10^{11}$ with transition from
the historic period to the modern one (see next Section), we can assume that just
significant changes of the dynamics of the inner layers of the
convection zone (most probably - of the bottom layer) were the main reasons 
for the transition from the historic to the modern period.

\section{Magnetic fields in the solar wind and on the Earth}

Although in the modern period the turbulent convection in the photosphere have no 
decisive impact on the sunspots {\it emergence}, the large-scale properties of the 
magnetic field coming through the sunspots into the photosphere and then to the 
interplanetary space (so-called solar wind) can be strongly affected by the photospheric motion. 
In order to be detected the characteristic {\it scaling} scales of this impact should be larger than 
the scaling scales of the interplanetary turbulence (cf. \cite{b1},\cite{gold}). In particular, 
one can expect that the cluster-exponent of the large-scale interplanetary magnetic 
field (if exists) should be close to $\alpha \simeq 0.37$. In Fig. 5 we show cluster-exponent 
of the large-scale fluctuations of the radial component $B_r$ of the interplanetary 
magnetic field. For computing this exponent, we have used the hourly averaged data obtained from
Advanced Composition Explorer (ACE) satellite magnetometers for the last solar cycle \cite{ACE}. 
As it was expected the cluster-exponent $\alpha \simeq 0.37 \pm 0.02$. Analogous result was 
obtained for other components of the interplanetary magnetic field as well. 

At the Earth itself the solar wind induced activity is measured by geomagnetic indexes such as aa-index 
(in units of 1 nT). This, the most widely used long-term geomagnetic index \cite{may}, presents long-term 
geomagnetic activity and it is produced using two observatories at nearly antipodal positions on 
the Earth's surface. The index is computed from the weighted average of the amplitude of the 
field variations at the two sites. We are interested in the clustering properties of the aa-index and 
their relation to the modulation produced by the photospheric convective motion. The point is that 
the data for the aa-index are available for both the modern and the historic periods. This 
allows us to check the suggestion that the Rayleigh number of the photospheric convection has the 
same order for the both mentioned periods. The transition between the two periods is clear seen 
in Figure 6, where we show the cumulated aa-index $a(t)$
$$
S_{aa} (t) = \int_{0}^{t} a(t') dt'  \eqno{(2)}
$$
The arrow in this figure indicates beginning of the transitional solar cycle.

Figure 7 shows cluster-exponents for both the modern and the historic 
periods calculated for the low intensity levels of the aa-index (for the historic period the level 
used in the calculations is aa-index=10nT, whereas for the modern period the level is aa-index=25nT, 
the aa-index was taken from \cite{wdc}). 
The low levels of intensity were chosen in order to avoid effect of the extreme phenomena (magnetic 
storms and etc.). The straight line in Fig. 7 is drawn to show the expected value of the cluster-exponent 
$\alpha \simeq 0.37$ for the both periods. It means that indeed for both the historic and the modern 
periods the Rayleigh number $Ra \sim 10^{11}$ (cf. Fig. 4) in the solar photosphere.  

\section{Chaotic sun}

The long-range reconstructions of the sunspot number fluctuations 
(see, for instance, Refs. \cite{sol},\cite{usos1}) allow us to look on the solar transitional 
dynamics from a more general point of view. In figure 8 we show a spectrum of such 
reconstruction for the last 11,000 years (the data, used for computation of the spectrum, 
is available at \cite{usos-recon}). The spectrum was computed using the maximum entropy method as in Ref. \cite{o} (see below). A semi-logarithmical representation was used in 
the figure to show an exponential law 
$$
E(f) \sim e^{-f/f_0}   \eqno{(3)}
$$
The straight line is drawn in Fig. 8 to indicate the exponential law Eq. (3). Slope of the 
straight line provides us with the characteristic time scale $\tau=1/f_0\simeq 176y (\pm 7y)$. 
The exponential decay of the spectrum excludes the possibility of random behavior and indicates the {\it chaotic} 
behavior of the time series. It is well known that low-order dynamic (deterministic) systems
have as a rule exponential decay of $E(f)$ in the chaotic regimes (see, for instance, \cite{o}). As for the delay-differential 
equations with chaotic attractors it is interesting to compare Fig. 8  with figure 3 of the Ref. \cite{fa}. 
It should be noted that the 176y period is the third doubling of the period 22y. 
The 22y period corresponds to the Sun's magnetic poles polarity switching. 

The exponential spectrum can be also 
produced by a series of Lorentzian pulses with the average width of the individual pulses equals to $\tau$ 
(though, the distribution of widths of the pulses should be fairly narrow to result in the 
exponential spectrum).

In Fig. 8 a local maximum corresponding to 
the frequency $f_0$ and its first harmonics have been indicated by arrows. It should be noted 
that the harmonic $2f_0$ corresponds to the well known period $T\simeq 88y$ \cite{fg}. Comparing this 
with the Fig. 1 one can conclude that the period (1933y-...) of the solar hyperactivity is close to its end.

\acknowledgments

The author is grateful to J.J. Niemela, to K.R. Sreenivasan, to C.
Tuniz, and to SIDC-team, World Data Center for the Sunspot Index,
Royal Observatory of Belgium for sharing their data and discussions.

\end{document}